# REAL-TIME PCR (QPCR) ASSAY FOR *RHIZOCTONIA SOLANI* ANASTOMOSES GROUP AG2-2 IIIB

SAYED JAFFAR ABBAS[1,2*], BASHIR AHMAD[1] AND PETR KARLOVSKY[2]

[1]*Center of Biotechnology and Microbiology, University of Peshawar, Peshawar, Pakistan*
[2]*Molecular Phytopathology and Mycotoxin Research Section, Georg-August-University Göttingen, Göttingen, Germany*
[*]*Corresponding author: e-mail: sayedjaffarabbas@hotmail.com*

## Abstract

*Rhizoctonia solani* anastomosis group AG2-2 IIIB is a severe sugar beet and maize pathogen. It causes crown and root rot disease which leads to yield losses world-wide. The soil-borne pathogen is difficult to detect and quantify by conventional methods. We developed a real-time PCR (qPCR) assay for the quantification of genomic DNA of *Rhizoctonia solani* AG2-2 IIIB based on the ITS region of rDNA genes. The limit of quantification of the assay is 1.8 pg genomic DNA. The amplification efficiency was 96.4. The assay will be helpful in the diagnoses of *Rhizoctonia solani* infection of sugar beet and maize roots and in the quantification of *R. solani* AG2-2 IIIB inoculum in plant debris and soil.

## Introduction

Genus Rhizoctonia comprisies a complex mixture of filamentous fungi with an imperfect state designated Rhizoctonia anamorph (*Rhizoctonia solani* Kühn). The anamorphic state does not produce conidia. The taxonomical classification of the Rhizoctonia anamorphs is based on hyphal fusion, dividing the complex into anastomosis groups (AG's) and sub groups (Sharon *et al.*, 2006). Disease symptoms caused by *R. solani* on host plants are also used as diagnostic features (Gonzalez & Rubio, 2006). It is sometimes difficult to accurately assign an isolate to an AG group because certain isolates do not anastomose with known representative of AG group or lose the ability of anastomoses (Hyakumachi & Ui, 1987), while some isolates anastomose with more than one anastomosis group (Sharon *et al.*, 2006).

*Rhizoctonia solani* is a soil-borne pathogen with a broad host range, causing diseases in a variety of crops, ornamentals and trees (Naz *et al.,* 2008, Gonzalez & Rubio 2006, Anderson 1982). In seedlings, *R. solani* causes damping off disease, producing black lesions in seeds and rot of plant parts which were in contact with soil during plant growth. Germination of basidiospores on the leaves surface results in foliar lesions. Fourteen anastomoses groups (AG's) have been described in *R. solani*, AG2 and AG4 being economically most important casual agents of root rot, damping off, hypocotyl rot, and fruit rot diseases (Harikrishnan & Yang, 2004). Based on the frequency of fusion between AG2 isolates, the group is divided into AG2-1 and AG2-2 subgroups. AG2-2 is further divided into subgroups AG2-2 IIIB and AG2-2 IV based on pathogenicity and morphology; the subgroups cannot be distinguished by anastomoses (Ogoshi, 1987).

*Rhizoctonia solani* AG2-2 IIIB causes Rhizoctonia root rot in sugar beet and maize plants, which leads to severe yield losses (Buddemeyer *et al.,* 2004, Strausbaugh *et al.,* 2011). In Germany, AG2-2 IIIB mainly causes roots and crown rot in sugar beet. No fungicides for the control of *R. solani* in sugar beets are approved in Europe (Buddemeyer *et al.,* 2004). The establishment of the pathogen was supported by narrow crop rotation (Ithurrart *et al.,* 2004). In maize AG2-2 IIIB causes round to elliptical, brown or black dry lesions on roots (Buddemeyer *et al.,* 2004). Depending on weather conditions (Ithurrart *et al.,* 2004), diseased plants may become completely rotten and lodged. The success of breeding maize for resistance against AG2-2 IIIB has so far been limited (Buddemeyer *et al.,* 2004).

Fungal nuclear rRNA genes are arranged as several hundreds of tandem repeats per genome. Each unit contains three genes: small rRNA genes 18S and 5.8S and the large rRNA gene 28S. These genes are often used in studies of fungal taxonomy and phylogeny. Conserved sequences exist in large subunits (LSU) and small subunits (SSU) while internal transcribed spacer (ITS) regions between the subunits are variable and therefore suitable for differentiation among closely related taxa.

Quantitative real-time PCR (qPCR) is suitable for taxon-specific quantification of pathogen DNA in infected host tissue or soil samples. Apart from delivering quantitative data, qPCR is faster and often more sensitive as compared to conventional PCR. qPCR for *Rhizoctonia cerealis* has been described by Guo *et al.*, 2012. In this study we developed a specific and sensitive qPCR assay for the quantification of *Rhizoctonia solani* AG2-2 IIIB in soil and plant samples.

## Material and Methods

**Fungal isolates and DNA extraction:** *Rhizoctonia solani* isolates used are listed in Table 1. Fungal DNA was extracted according to a CTAB protocol (Brandfass & Karlovsky, 2006).

**Primer selection and specificity test:** A number of forward and reverse primer combinations were tested. Forward primers selected from ITS1 18S region were Rhsp1 (AACAAGGTTTCCGTAGGTG), AG2sp (ATTATTGAATTTAAACAAAG), and AG22sp2 (TAGCTGGATCCATTAGTTTG) (Salazar *et al.*, 2000). These primers were combined with the reverse primer 5.8SKhotR (GTTCAAAGAYTCGATGATTCAC) (Fredricks *et al.*, 2010), amplifying a product of 322 to 330bp. The qPCR system with SYBR Green detection was optimized for 25μL reactions containing PCR buffer (Bioline, Germany), 3mM $MgCl_2$, 200μM dNTPs (Bioline, Germany), 0.3μM forward primer, 0.3μM reverse primer, 10nM fluorescein, 0.1x of SYBR Green 1:1000 (Molecular Probes, USA), 0.25U of Taq DNA polymerase (Bioline, Germany) and 1μL of template DNA. iCycler (BioRAD, Life Sciences Group, California, USA) was used with the following temperature profile: initial denaturation 94°C/3 min was followed by 40 cycles of 94°C/30s, 50-60°C gradient/20s and 72°C/30s and by a final extension

72°C/5 min.

Table 1. Fungal strains.

| Fungal species | Isolate | Anast. group | Source | Fungal species | Isolate | Source |
|---|---|---|---|---|---|---|
| *Rhizoctonia solani* | MP93 | AG2-2 IIIB | 5 | *Fusarium equiseti* | FE 5 | 5 |
| *Rhizoctonia solani* | Tester | AG1-IB | 1 | *Fusarium cerealis* | FCKW 3 | 5 |
| *Rhizoctonia solani* | Tester | AG2-1 | 1 | *Fusarium cerealis* | FCKW 4 | 5 |
| *Rhizoctonia solani* | Tester | AG3-PT | 1 | *Fusarium sacchari* | FSAC 1 | 3 |
| *Rhizoctonia solani* | Tester | AG4-HGII | 1 | *Fusarium sacchari* | FSAC 2 | 3 |
| *Rhizoctonia solani* | GC | AG5 | 1 | *Fusarium culmorum* | FC 15SS | 5 |
| *Rhizoctonia solani* | PL | AG5 | 1 | *Fusarium culmorum* | FC 22SS | 5 |
| *Rhizoctonia solani* | PL-1047 | AG5 | 1 | *Fusarium culmorum* | FC 2 | 5 |
| *Rhizoctonia solani* | GER | AG5 | 1 | *Fusarium culmorum* | FC H15 | 5 |
| *Rhizoctonia solani* | GER 1042 | AG5 | 1 | *Fusarium culmorum* | FC H69 | 5 |
| *Rhizoctonia solani* | Tester | AG8 | 1 | *Fusarium culmorum* | FC H71 | 5 |
| *Rhizoctonia solani* | 5P | AG8 | 1 | *Fusarium culmorum* | FC H73 | 5 |
| *Rhizoctonia solani* | GB | AG11 | 1 | *Fusarium verticillioides* | FV 3 | 3 |
| *Rhizoctonia solani* | Tester | AG11 | 1 | *Fusarium verticillioides* | FVER 420 | 5 |
| *Rhizoctonia solani* | 1079 | AGD | 1 | *Fusarium verticillioides* | FVER 429 | 5 |
| *Rhizoctonia solani* | Tester | AGE | 1 | *Fusarium proliferatum* | FPROL 12 | 5 |
| *Fusarium oxysporum* | FO 2 | | 2 | *Fusarium proliferatum* | FPROL 1 | 3 |
| *Fusarium oxysporum* | FO | | 5 | *Fusarium proliferatum* | FPROL 2 | 3 |
| *Fusarium oxysporum* | FO 125 | | 5 | *Fusarium proliferatum* | FPROL 4 | 3 |
| *Fusarium oxysporum* | FO 436 | | 5 | *Fusarium proliferatum* | FPROL 5 | 3 |
| *Fusarium oxysporum* | FO 121SS | | 5 | *Fusarium proliferatum* | FPROL 6 | 5 |
| *Fusarium avenaceum* | FA 3 | | 5 | *Fusarium proliferatum* | FPROL 7 | 5 |
| *Fusarium avenaceum* | FA 9 | | 5 | *Fusarium proliferatum* | FPROL 8 | 5 |
| *Fusarium avenaceum* | FA 1.2 | | 3 | *Fusarium proliferatum* | FPROL 9 | 5 |
| *Fusarium avenaceum* | FA 5.2 | | 5 | *Fusarium proliferatum* | FPROL 11 | 5 |
| *Fusarium acuminatum* | FACU 1 | | 4 | *Fusarium graminearum* | FGR 62048 | 3 |
| *Fusarium acuminatum* | FACU 3 | | 5 | *Fusarium graminearum* | FGR 62722 | 3 |
| *Fusarium acuminatum* | FACU 5 | | 5 | *Fusarium graminearum* | FGR 67638 | 3 |
| *Fusarium equiseti* | FE 2 | | 4 | *Fusarium graminearum* | FGR 83649 | 3 |

1- Strain collection of Division of general plant pathology and crop protection, Georg-August-University, Göttingen, Germany
2- Mykothek FAP (Dr. M. Winter)
3- Deutsche Sammlung von Mikroorganismen und Zellkulturen, Braunschweig, Germany
4- International Center for Agricultural Research in the Dry Areas, Aleppo, Syria
5- Strain collection of the Molecular phytopathology and mycotoxin research division, Georg-August-University, Göttingen, Germany

**Sequence analysis:** PCR product obtained from *R. solani* AG2-2 IIIB (strain MP93) was directly sequenced by using primers used for the amplification (MWG Biotech, Munich, Germany).

**Calibration and determination of PCR efficiency:** To determine the sensitivity of the assay, *R. solani* AG2-2 IIIB (isolate MP93) genomic DNA was quantified by densitometry as described by Nutz *et al.*, 2011 using ImageJ software (http://r sbweb.nih.gov/ij/). Calibration curves were constructed by plotting threshold cycle value versus DNA concentration (3-fold dilution series from 444.4 pg to 1.8 pg). After the PCR, melting curves of the products were recorded using the same thermocycler.

### Results

On the basis of preliminary experiments with primer combinations described in the Material and Methods section, forward primer AG22sp2 in combination with reverse primer 5.8SKhotR were selected. qPCR conditions for the primer pair were optimized, leading to the temperature profile specified in Material and Methods section.

The sequence of the forward primer AG22sp2 was compared with sequences in NCBI database using BLAST. All hits with 100% sequen ce identity originated from *Rhizoctonia solani* isolates belonging to anastomoses groups AG-2, AG2-2 IIIB and AG-2-2 IV. The sequence of the reverse primer was no t tested because it was expected to match most fungal species. Amplicon generated with primer pair AG22sp2/5.8SKhotR from genomic DNA of *R. solani* isolate MR93, which belong s to anastomosis group AG2-2 IIIB, was sequenced (NCBI accession number JX914627). Comparison with the database the seq uence was found identical with sequences from *Rhizoctonia solani* anastomose group AG2-2 IIIB isolates (Acc. Nos. GU811684.1, FJ492151.3, FJ492138.3, FJ492137.3, FJ492136.3, FJ492124.3, FJ492123.3, and FJ492089.3.). To extend the characterization of the specificity of primer AG22sp2 described by Salazar *et al.,* 2000 who desi gned the primer (Salazar *et al.,* 2000), genomic DNA of *R. solani* isolates from anastomosis groups AG1-IB, AG2-1, AG3-PT, AG4-HGII, AG5, AG8, AG11, AGD, AGE and DNA of common fungal pathogens of m aize and sugar beet *Fusarium oxysporum*, *F. a venaceum*, *F. acuminatum*, *F. equiseti*, *F. cerealis*, *F. sacchari*, *F. culmorum*, *F. verticillioides*, *F. proliferatum* and *F.graminearum* (Table 1) were tested in the qPCR assay. The results were negative for

*R. solani* isolates AG1-IB, AG2-1, AG3-PT, AG8, AG11, AGE and AG4-HGII (all isolated from the set "Tester"), AG5 (isolate GC), AG5 (isolates PL), AG5 (isolate PL-1047), AG5 (isolate Ger), AG5 (isolate Ger 1042), AG8 (isolate 5P), AG11 (isolate GB), and AGD (isolate 1079). The assay was also negative for DNAs extracted from fungal species other than *R. solani* except for *Fusarium avenaceum* FA5-2 and *Fusarium graminearum* FGR83649, which produced amplicons late in the PCR with threshold cycles larger than 32. DNA of the other isolates of *F. avenaceum* and *F. graminearum* (Table 1) tested negative.

Linear relationship between the logarithm of DNA concentrations and the threshold cycle was found in the range of 1.8 pg to 444.4 pg DNA with a coefficient of determination $R^2 = 0.99$. The amplification efficiency with pure template DNA was 96.4% (Fig. 1). Melting temperature of the products was 86°C (Fig. 2).

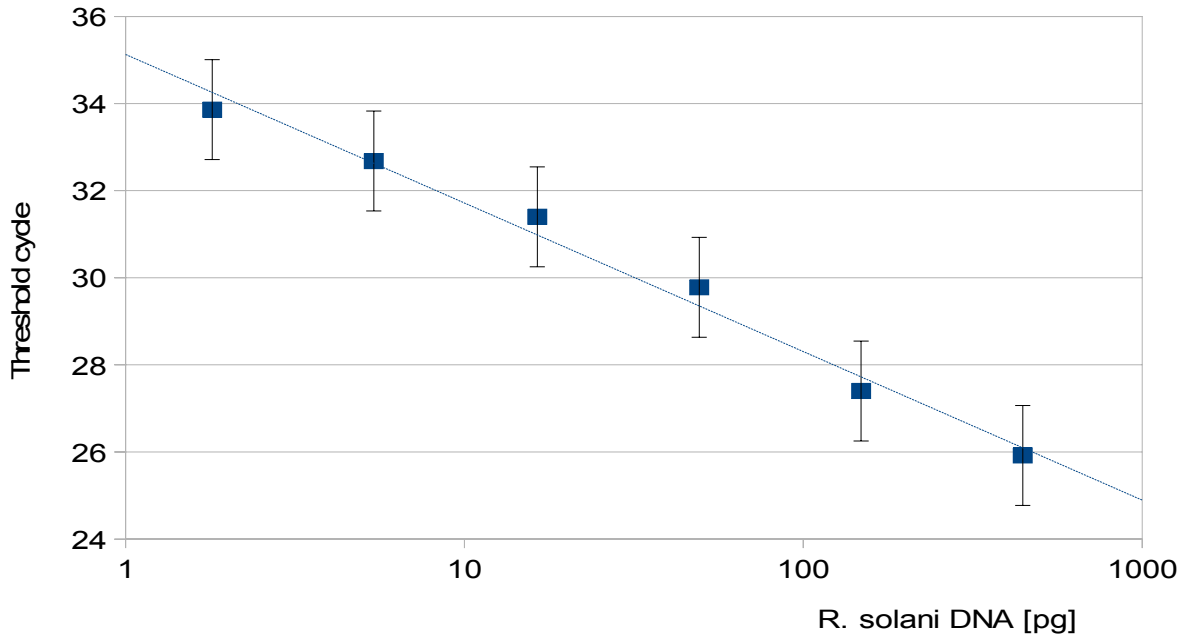

Fig. 1. Standard curve for qPCR for genomic DNA of *Rhizoctonia solani* AG2-2 IIIB. DNA standard series from 444.4 pg to 1.8 pg was used; whiskers show standard deviation.

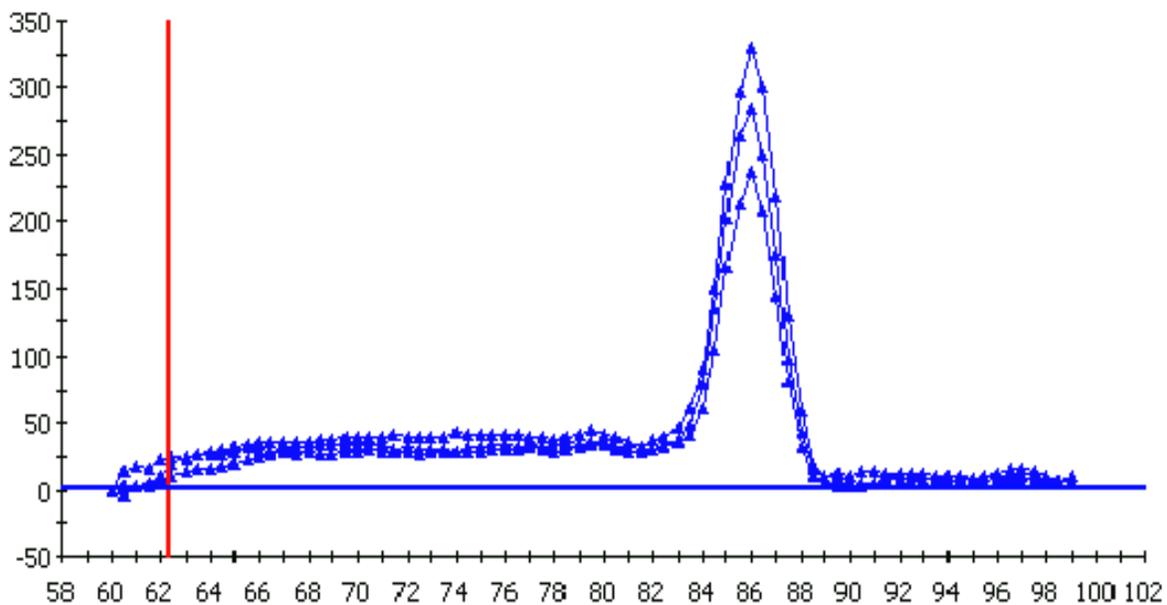

Fig. 2. Melting curves of amplification products of *Rhizoctonia solani* AG2-2 IIIB DNA at three concentrations. The melting temperature was 86°C.

## Discussion

Detection of *Rhizoctonia solani* anastomoses groups AG2-2 IIIB, AG-2, and AG-2-2 IV by conventional PCR-RFLP has been reported using ITS1 and ITS4 primers (Hyakumachi *et al.*, 1998). These primers produced an amplicon of 740bp, which is too large for qPCR. Detection of *Rhizoctonia solani* AG2-2 with conventional PCR was also reported by Matsumoto (Matsumoto, 2002) but the primers developed in this work amplified isolates of AG5 and AG7, too. The assay reported in this communication is thus the first qPCR assay for *R. solani* anastomosis group AG2-2 IIIB.

## Conclusion

We envision that the assay will be useful in the diagnoses of *R. solani* infection of sugar beet and maize and in the quantification of the inoculum of the pathogen in plant residues and soil.

## Acknowledgments

The study was supported by the Ministry for Science and Culture of Lower Saxony within the network KLIFF-climate impact and adaptation research in Lower Saxony.